\documentclass[12pt]{article}
 \usepackage{amsfonts}
 \usepackage{amssymb,amsmath}
 \usepackage{graphicx} \usepackage{epstopdf}
 \newcommand{\crlb}[1]{\label{#1}\\[2pt]}
 
 \newcommand{\crld}[1]{\label{#1}}
 \newcommand{\eela}[1]{\quad\hbox{\scriptsize{#1}}\label{#1}\end{eqnarray}}
 \newcommand{\eelb}[1]{\label{#1}\end{eqnarray}}
 
 \newcommand{\newsecb}[2]{\section{#1}\label{#2}\setcounter{equation}{0}}

 \newcommand{\nolabels} {\def\eel{\eelb}\def\eeql{\eeqlb}  \def\crl{\crlb} 
 \def\newsecl{\newsecb}\def\bibiteml{\bibitem} \def\citel{\cite}\def\labell{\crld}}
\newcommand{\eeqla}[1]{\quad\hbox{\scriptsize{#1}}\label{#1}\end{aligned}\end{equation}}
\newcommand{\eeqlb}[1]{\label{#1}\end{aligned}\end{equation}}

\newcommand\publishversion  {\nolabels\setlength{\textheight}{8.38in}\setlength
    {\oddsidemargin}{0in} \setlength{\textwidth}{6.2in}\setlength{\topmargin}{-0.2in}}
\def\beq{\begin{equation}\begin{aligned}}		\def\eeq{\end{aligned}\end{equation}}
\def\be{\begin{eqnarray}}  					\def\ee{\end{eqnarray}}		
   \def\bi#1{\begin{itemize}\item[#1]} 	      	   \def\ei{\end{itemize}} 
   \def\eqn#1{(\ref{#1})}
   	 \def\fn{\footnote}	 

		 \def\del{\delta}  % \def\k{\kappa}     \def\l{\lambda}  
    		  	\def\D{\Delta}    
	    		        		     	\def\vv{\varphi}     
 	 		\def\s{\sigma}     	      	 
	     		          		              	\def\w{\omega}  
    		  		 		\def\HH{{\mathcal H}}

          		\def\ra{\rightarrow}	
\def\bra{\langle} 		\def\ket{\rangle}

\def\fract#1#2{{\textstyle\frac{#1}{#2}}}	 	 	
\def\ffract#1#2{\raise .2 em\hbox{$\scriptstyle#1\,$}\kern-.34 em/\kern-.34 em\lower .15 em \hbox{$\scriptstyle\,#2$}}

\def\ex#1{e^{\textstyle#1}} 			
\def\bpmatrix{\begin{pmatrix}} 			\def\epmatrix{\end{pmatrix}}
\def\bmatrix{\begin{matrix}} 			\def\ematrix{\end{matrix}} 
\def\bcenter{\begin{center}}			\def\ecenter{\end{center}}

\def\lowerheightfig#1#2#3{\(\raise-#1\hbox{\includegraphics[height=#2]{#3}}\)}
\def\lowerwidthfig#1#2#3{\(\raise-#1\hbox{\includegraphics[width=#2]{#3}}\)}

\def\intt{{\mathrm{int}}}    
 \def\tot{{\mathrm{tot}}}

\def\weglaten#1{}	
 
 \def\twomat#1#2{\Big(\begin{matrix} #1 \\ #2 \end{matrix}\Big)} %\twomat{a&b&c}{&d&} etc.
 \publishversion
 
\begin{document}
\begin{titlepage} 
\title{
 Explicit construction of Local Hidden Variables for any quantum theory up to any desired accuracy\\[5pt]
\normalsize Version 3}
\author{Gerard 't~Hooft}
\date{\normalsize
Faculty of Science,
Department of Physics\\
Institute for Theoretical Physics\\
Princetonplein 5,
3584 CC Utrecht \\
\underline{The Netherlands} \\[10pt]
e-mail:  g.thooft@uu.nl \\ internet: 
http://www.staff.science.uu.nl/\~{}hooft101
}
 \maketitle

\begin{quotation} \noindent {\large\bf Abstract } \\[10pt]
The machinery of quantum mechanics is fully capable of describing a single realistic world.  Here we discuss the converse: 
in spite of appearances, and indeed numerous claims to the contrary, any quantum mechanical model can be mimicked, up to any finite accuracy, by a completely classical system of equations. An implication of this observation is that Bell's theorem is not applicable in the cases considered. This is explained by scrutinising Bell's assumptions concerning causality, retrocausality, statistical (in-)dependence, and his fear of `conspiracy' (there is no conspiracy in the language used to describe the deterministic models). The most crucial mechanism for the counter intuitive Bell/CHSH violation is the fact that, regardless the settings chosen by Alice and Bob, the initial state of the system should be a realistic one.  The potential importance of our construction in model building is discussed.		% 150 ww
 \end{quotation}\end{titlepage}
	
\newsecl{Introduction}{intro}
Quantum mechanics is usually perceived as being a revolutionary new theory for the interactions and dynamics of tiny particles and the forces between them. Here we expand on our earlier proposal\,\cite{GtHCA.ref} to look at quantum mechanics in a somewhat different way. The fundamental interactions could be entirely deterministic, taking place in a world where all laws are absolute, without requiring statistics to understand what happens. The infinite linear vector space called Hilbert space, is then nothing more than a mathematical utensil, allowing us to perform unitary transformations. The set of all possible states is re-arranged into states that look like superpositions of the original ontological, or realist,\fn{Everywhere in this paper, the words `ontological' and 
`realist' are used interchangeably. They emphasise that we avoid `statistical' or `uncertain' expressions.}
states. As soon as we loose control of these original realist states, we can, instead, interpret the absolute values squared of the superposition coefficients as Born probabilities. Quantum mechanics as we are familiar with today, is then arrived at.

This paper is about regaining control. The `original, realist states' then are what is referred to as `hidden variables'. Let there be given some quantum model, for instance the Standard Model of the fundamental particles and their interactions. Can we then identify these hidden variables? Often, hidden variables are looked upon as rather ugly attempts to regain a dated interpretation of quantum theory.  In contrast, this theory can be far more beautiful and elegant than the original theory of quantum mechanics as displayed in the Copenhagen frame of axioms. This is what we intend to show in this paper. We start with a generic ontological theory. This may seem to be hard to visualise, until we realise that demanding locality severely restricts the possibilities, see Section~\ref{weird.sec}. One can almost derive what the hidden variables are: Local Hidden Variables (LHV).

It was thought that local hidden variables should be easy to refute, but this, we now claim, is a mistake. The 
use of quantum mechanics as a procedure for \emph{vector analysis of a classical system}, is merely a mathematical trick, which does not change the equations;  therefore it does not introduce `conspiracy' either in the classical or in the quantum description.

In this paper, we first explain the use of quantum mechanics to describe deterministic systems in their full generality (section \ref{generalonto.sec}). Basically, a system is deterministic if its evolution law can be regarded as an element of a large permutation group. We explain how it can be mapped on the set of unitary transformation matrices in Hilbert space. There are various advantages of using such representations in pure mathematics. The Schr\"odinger equation appears naturally, reflecting what happens if one uses the fact that unitary matrices form a continuum while permutations are discrete. In short: any deterministic theory can be written as a quantum theory, in the sense that one may use the concept of Hilbert space for doing statistics, and a  Schr\"odinger equation describes the evolution law.

Next, in section \ref{ontomodel.sec}, we show the converse of that: how any quantum system can be linked to a deterministic model. The observation we use is that, in any classical or quantum system, one may add physical degrees of freedom that move periodically in a compact space. This adds new energy levels to the system that we assume to be invisible, in particular if the detection devices used have a limited time resolution. We call these `fast fluctuating variables'. Our new twist is that this compact space may be assumed to be sufficiently tiny, so that these variables return to their previous values with very high frequencies. Consequently, in our artificial vector spaces, their energy spectrum is discrete with wide separations between the energy levels. These separations are so large that, under normal physical conditions, only the very lowest energy state will be occupied. This state is a single wave function, which happens to be a constant: \emph{Our fast variables are all in uniform probabilistic distributions.} This makes them invisible in practice, just as the fast vacuum fluctuations due to the fields of extremely heavy virtual particles.

Nevertheless, we now have quantum states that may interact with the slow particles non-trivially.
If now we construct the \emph{vector space representation} of such a model, we shall find that any quantum model one wishes to investigate can emerge from this.

Some readers noted that our model is non-local. This was due to an error in the notation that could easily be corrected. In Section \ref{nonlocal.sec}, we explain the situation. Locality is not a problem, but special relativity is. This is due to the fact that we need to restrict ourselves to finite models, forcing us to work with lattice theories. Making such theories relativistically invariant is notoriously difficult. It will presumably involve general relativity, and this is far beyond what we can handle presently.

Of course, the reader may wonder how it can happen in such a model that Bell's theorem\,\cite{Bell.ref}--\cite{MaryBell.ref} is disobeyed. How could this happen? 
This we describe in section \ref{bell.sec}. Actually, we have to deal with two questions: one, what was wrong or misunderstood in Bell's classical argument, and two, what is the origin of the apparent clash between totally natural intuitions on the one hand, and the actual quantum calculation -- as well as the real experiment -- on the other. We think both questions can be answered, but we keep the answers brief. Thus we summarise what, according to this author, the principal weaknesses are in Bell's argument, which is not the mathematical calculations but the general assumptions, in particular those connected with causality and `free will'. The question why our intuition does not agree with the result of the experiments is actually more interesting. It all comes from an important footnote in our theory: the realistic degrees of freedom depend on the basis chosen; one can constrain the choice chosen by nature by imposing more demands. Of our various options for the cause of our discrepancy with Bell, the last explanation given in Section \ref{bell.sec}, `option \# 4', appears to be the salient one. It explains why and how the `free will' of Alice and Bob to change their settings, enters into the argument: they do \emph{not} have the free will to choose their settings in some quantum superimposed state, since the initial state of the universe is a realistic one.

There are numerous examples of quite counter intuitive consequences of the original Copenhagen interpretation. All these can be traced to the same features. We mention a few in section \ref{weird.sec}: the reason why quantum mechanics is weird, arises from the fact that deterministic underlying theories do not work in the way expected. The deterministic variables do randomise when generating their `quantum effects', but this goes along mathematical paths that differ from our intuitive ideas. Their behavior looks as if `conspiracies' take place, while these variables do nothing else than obeying physical laws.

We claim that this paper is not just a philosophical one, but to the contrary, it may well show novel guidelines towards building models, see section \ref{model.sec}. Our coclusions are in Section~\ref{conc.sec}.

\newsecl{The generic realistic system}{generalonto.sec}
Let us briefly review the precedure described in Ref.\,\cite{GtHCA.ref}.
In a deterministic theory, we identify the set of states that can be realised. In general, the number of different possible states, \(N\), will be gigantically large,  but the principle will always be the same. The dynamics is defined by the evolution law over a smal lapse of time, \(\del t\). This is nothing but a single element of the permutation group \(S_N\):
\be |k(t+\del t)\ket=U(\del t)|k(t)\ket\ ;\qquad U(\del t)\in S_N\ .\ee
 It maps the set of states onto itself. This set should be regarded as discrete, although it will often be useful to consider one of several possible continuum limits, so as to simplify notations and calculations.

It will be useful to regard our system as bounded, by imagining it to be surrounded by a wall. Later, the wall will be removed, enabling us, for instance, to set up plane wave expansions. We start with choosing the time variable as being discrete, \(t=k\, \del t\), where \(k\) is integer and \(0\le k<N\).

\(N\) is finite. This implies that there is a  number \(T\le N\) such that
\be |k(t+T)\ket=|k(t)\ket\ ;\qquad |k(t+t_1)\ket \ne |k(t)\ket \ \  \hbox{ if } \ \ 0<  t_1< T\ . \eel{cycle.eq}
Thus, our system is periodic in time (though the period \(T\) rapidly tends to infinity as we choose our wall to be further away).
In general \(T\ll N\), so that, consequently, there will be many states \( |k\ket\) that are not in the periodic set \eqn{cycle.eq} at all. Starting with such a state gives us another periodic set with period \(T'\). Continuing this way, we find that the completely generic finite deterministic system consists of a large number of periodic sets.

Each periodic set \((r)\) can now be subject to a discrete Fourier transform, to write its elements as superpositions of energy eigenstates \(|n,r\ket\,,\ n=0,\cdots T_r-1\,,\)
\be |k(t),r\ket=\frac 1{\sqrt T_r}\sum_{n=0}^{T_r-1}\ex{-iE_{n,r} t}|n,r\ket\ ,\qquad E_{n,r}=\frac{2\pi n}{T_r\,\del t}+\del E_r\ ,
\eel{energysequence.eq}
where \(\del E_r\) is, as yet, an arbitrary normalisation of the energy that may well be different for different sets  \(r\). Its physical interpretation is that the energy may depend in an arbitrary way on the quantum number \(r\), which obeys a conservation law. Note, that the concept of energy \(E\) used here is the quantum mechanical one, not necessarily the classical Hamiltonian.

The above describes the formal, complete solution of all deterministic systems.
In a given set \(r\), the energy spectrum \eqn{energysequence.eq} forms an equidistant sequence. If we take the limit where \(\del t\) vanishes while \(T_r\del t\) is kept fixed, this spectrum ranges to infinity. At given \(r\), the distance between adjacent energy eigenvalues, \(\D E=E_{n,r}-E_{n-1,r}\)\,,  is determined by the period,  \(\D E=2\pi/(T_r\,\del t)\). They are strictly equidistant, regardless how complicated the interactions may be.

Since the energy eigenvalues do not depend on the basis chosen, the equidistant energy level sequences make one wonder how well such a theory can represent the real world, where such exactly equidistant sequences seem not to be observed. The answer to this will be that there are equidistant sequences, but the energy separations are too wide to be noticeable in physics. We'll see how this can happen.

\newsecl{A deterministic theory for every quantum model}{ontomodel.sec}
In Ref.~\cite{GtH-2020.ref}, it is found how a deterministic model can be constructed that mimics any given quantum system. Here we briefly repeat the derivation, by first formulating the deterministic model, and then showing how to link its dynamical evolution law to the given Schr\"odinger equation. We shall then explain why the result is counter intuitive.

For simplicity our quantum system will be described as a one-particle theory, but generalisation to more complex structures such as quantum field theories will be straightforward.

The classical  model contains \(N\) primary states, \(|i\ket,\ i=1,\cdots,N\), which we declare all to be realistic. These will later be declared to correspond to the basis elements of the quantum model,  but we'll come to that. In addition to these primary states, there will be a number,\fn{Our notation often changes in different publications.} \(M\), of \emph{fast fluctuating, periodic variables} \(\vv_i(t)\,, \ \ 0\le\vv_i<2\pi\), where  \(i=1,\cdots, M\).

Anticipating demands that will be needed later (in arguments concerning locality, see Section \ref{nonlocal.sec}), we shall choose \(M\) to be the number of points in 3-space, regardless the other properties of the system under investigation: 
\be M=\#(\vec x\,)\ . \eel{numberofpoints.eq}
What these fast variables actually are, is to a large extent immaterial, as long as they move so fast that all positions are taken more frequently than the largest quantum frequencies in the system we wish to describe. For instance,  we may suggest that the variables might represent excessively heavy virtual particles, situated at the space points \(i\in M\).

In view of the above, the evolution law of \(\vv_i\) is written as
\be\vv_i(t)=\vv_i(0)+\w_i t\mod 2\pi\ ,\qquad  \ t=k\,\del t\,, \eel{phievolve.eq}  
where \(k\) is integer while \(\del t\) is very small (though not infinitesimally small).
The periods \(T_i\) are given by  integers \(L_i\), such that \(T_i=L_i\,\del t\,,\) and \(\w_i=2\pi/T_i={2\pi}/{(\del t\,L_i)}\). This means that the variables \(\vv_i\) are arranged to sit on an \(M\)-dimensional lattice with \(\prod_{i=1}^M L_i\) points, and periodic boundary conditions in all \(M\) directions. The values for the \(\w_i\) are large compared to the physical time parameters, so  \(\del t\) and \(L_i\,\del t\) are all small, but \(L_i^2\,\del t\) will be assumed to be relatively large.

We choose the \(L_i\) large but not exactly equal. It will make things easier if we assume them to be relative primes (so that the \(\vv_i\) will mix properly).

The primary states \(|i\ket\), also called `slow variables' in Ref.~\cite{GtH-2020.ref}, undergo deterministic transitions whenever the variables \(\vv_i\) reach some given points in their lattice. These transitions are all elements of the permutation group \(S_N\).

For simplicity, we limit ourselves to pairwise permutations between two states at the time. In general, we assume for each pair \((i,j\)) a number \(n_{i,j}\) of given, fixed points in the 2 dimensional sub-lattice spanned by \(\vv_i\) and \(\vv_j\), such that, if \(\vv_i\) and \(\vv_j\) arrive simultaneously at one of the points \(n_{i,j}\), the transition \(|i\ket\leftrightarrow |j\ket\) takes place. The \(\vv\) variables just continue obeying \eqn{phievolve.eq}. Let us call these special points on the \(\vec\vv\)-lattice \emph{crossing points}.

When locality is considered, we choose  \(i\) and \(j\) to be neighbors in 3-space.

So-far, our model is just like any other realistic, deterministic model. But
now comes the most important constraint: \emph{the \(\vv_i\) variables go so fast that we cannot detect their values directly.} Therefore, we assume them to be in a totally even probabilistic distribution in the entire space spanned by them all. In contrast, the primary variable hops from one state to others at a slow pace, since it happens relatively infrequently that the \(\vv\) variables 
arrive simultaneously at one of the crossing pints. When we compute what happens classically, we see that the distribution of the fast variables stays even, but how do the states \(|i\ket\) behave? This is most easily found out by using the quantum mechanical notation, as described in the previous section. It was also discussed in  ref.~\cite{GtH-2020.ref}. Simply re-write the time evolution described above as an equation for a deterministic, \emph{real} wave function \(\psi_i(\vec \vv,\,t)\). The dominant part of the Hamiltonian takes care of Eq.~\eqn{phievolve.eq}. Then, we have perturbation parts for each pair \((i,j)\), taking the form
	\be H_{ij}^\intt=\sum_{i,j,s}\frac{\pi} 2\s_y^{[i,j]}\,\delta(k_i=k_i^{(s)})\,\delta(k_j=k^{(s)}_j)\ ,\eel{flipH.eq}
where \(k_{i}^{(s)}\) and \(k_j^{(s)}\) indicate the crossing point(s) on the \((i,j)\) sub-lattice. Here, the number \(\del t\) was set equal to one for simplicity, and `int' stands for `interaction'.

The pairwise permutation between the states \(|i\ket\) and \(|j\ket\) is written as a Pauli matrix,
	\be -i\s_y^{[i,j]}=\twomat{0&-1}{1&0} \ ,\eel{paulimaty.eq}

We regard the total Hamiltonian describing the transitions between the fast and the slow variables,  over small time steps \(\del t\) or \(L\,\del t\), as  small perturbations. We have to assume that the fast variables are in their lowest energy state. Since their excited energy levels are far separated form the lowest value (zero) and energy is exactly conserved, this situation is stable and therefore persists in time. Keeping them in the lowest energy state guarantees that the statistical distribution on the \(\vv\) lattice stays perfectly even as time proceeds.\fn{The matrix elements \(\bra\vec\phi\,|E_r\ket\) may be assumed to have exactly the same amplitude everywhere on the hypertorus of the \(\phi\) lattice.}

Since we consider small steps in time, perturbation theory tells us that this is controlled by the \emph{expectation value} of the Hamiltonian \(H_{ij}^\intt\) for the zero energy states of the \(\vec\vv\) variables. This replaces the Kronecker deltas in Eq.~\eqn{flipH.eq} by the factor
	\be \frac1 {L_iL_j}\ . \eel{fracfactor.eq}

This is very important. The expression in Eq.~\eqn{flipH.eq}, with  the factor \(\pi/2\) included, describes a completely deterministic exchange of states, as it was designed to do. But the factor \eqn{fracfactor.eq} turns this into a Hamiltonian that generates superpositions. Thus, from here on, we are dealing with real quantum mechanics.

The outcome of our calculation in perturbation theory is, that the probability distribution of the primary state, can be written as 
	\be P_i(\vec\vv,t)=P_i(t)=\psi_i(t)^2\ , \eel{slowprob.eq}
where \(\psi_i(t)\) form a \emph{real-valued} wave function. This wave function slowly varies in time, as we already explained.
The \(\vv_i\) run along their respective circles so quickly that the even distribution of the probabilities, given by the norm \(\sum_{i=1}^N|\psi_i|^2\), does not show any significant  dependence on the variables \(\vv_i\) or on time itself. This takes care of unitarity in the space of the slow variables. 

The Hamiltonian we obtain is
	\be H_{\mathrm{slow}}=\sum_{i,j}\frac{\pi\,n_{i,j}} {2L_iL_j}\s_y^{[i,j]}\ ,\eel{Hslow.eq}
where \(n_{i,j}\) is the number of crossing points on the  (2-dimensional) lattice spanned by the values for \(\vv_i\) and \(\vv_j\).\fn{It is here that we prefer to  assume \(L_i\) and \(L_j\) to be relative primes, so as to ensure that any crossing point will be passed equally often.}

This is a purely imaginary, antisymmetric matrix. Such Hamiltonians keep the wave function real, and have no diagonal parts. By choosing the numbers \(n_{i,j}\,, L_i\,,\) and \(L_j\) we can generate almost any such antisymmetric, imaginary-valued Hamiltonian. If the desired Hamiltonian has terms with irrational ratios in them, the limits \(n,\,L\ra\infty\) have to be taken.

So-far, the wave function here came out to be real. Usually, in quantum mechanics, we have complex valued wave functions. This is easy to arrange here. Complex numbers are pairs of real numbers, so having complex numbers means that there is an extra, somewhat hidden, binary degree of freedom, called `c-bit' in ref.~\cite{GtH-2020.ref}, such that this c-bit takes the value 0 for the real part of the wave function, and 1 for the imaginary part. It was easier for us to start with the real numbers only. The c-bit doubles the total number of slow states.

Comparing the quantum calculation with what happens classically, we can see what exactly is happening. Quantum mechanically, a wave function \(\psi_i(t)\) obeys a Schr\"o\-dinger equation. Classically, the system undergoes transitions from the states \(|i\ket\) to all other states \(|j\ket\) in fairly rapid successions. Our classical theory now tells us what is causing these transitions: they depend on the exact values of the \(\vv\) variables, which however are difficult to follow since they go fast. Only if measurements are done so fast that we know exactly where the \(\vv\) variables are, the evolution will turn out to be deterministic. But as long as we cannot follow the fast variables, we have to deal with the quantum expressions.

The calculation of the effective Hamiltonian for the slow states may require higher order corrections if the numbers \(n,\,L,\,\dots\) are not very large. This has not been done yet, partly because it is the principle that counts here. In higher order corrections, one may have to deal with some of the higher energy states of the \(\vv\) variables as virtual states; the emergence of virtual \(\vec\vv\) states at higher energy levels then betrays deviations from the strictly even distribution of probabilities for \(\vec\vv\). Much like the effects of heavy, virtual particles, these higher order effects may represent virtual particles that cause non-local interactions and other complications. This would not lead to any violation of quantum mechanics itself, in spite of the deterministic underlying theory.
 
 \newsecl{Special features. Locality}{nonlocal.sec}

There may seem to be some arbitrariness in the above construction. Why do transitions between the states \(|i\ket\) and \(|j\ket\) only involve the fast variables \(\vv_i\) and \(\vv_j\)? Why not the others? Also, the question may arise: where do we have to choose the crossing points, and how does the theory depend on this choice? And then: why do we not unfold the \(M\)-dimensional lattice of all \(\vv\)-variables to form just one line of consecutive states, with length \(L_\tot=\prod_iL_i\)? 

All these questions have to do with locality. We may wish to avoid theories where the properties of a particle at site \(\vec x\) depend on things happening at site \(\vec y\) far away from \(\vec x\). In particular, in second-quantised theories this may be important. Signals should not go faster than light. Classically, such conditions are easily seen to be obeyed if the crossing points \((i,j)\) can be associated with one single position in real space (or space-time). Opening up the \(\vec\vv\) lattice would generate difficulties.
It would force us to re-arrange the energy levels of the \(\vv\) variables to form much larger sequences with much smaller energy gaps;  we would not be able anymore to keep the higher energy levels out of the discussion on how real quantum mechanics can be obtained.
	
	Thus we see that demanding locality narrows down much of the arbitrariness of our models, though some arbitrariness may seem to remain, such as the choice of the locations of the crossing points on the lattices. However, for these points, another remark may be made. We can handle the effects of the crossing points by re-arranging the field components in terms of eigenstates of the permutation operator considered. We could consider the wave functions \(\psi_i\) and  \(\psi_j\) and decompose them along the lines of the eigenstates,
		\be \psi_\pm=\fract 1{\sqrt 2}(\psi_i\pm\psi_j)\ , \eel{plusminus.eq}
We then find that, at a crossing point, \(\psi_+\) is not affected at all, while \(\psi_-\) gets a minus sign. but this minus sign could be chosen differently, which would amount to displacing the crossing point. Thus we obtain a symmetry of the system, implying that the exact location of the crossing point is, to some extent, physically immaterial.

% xxxx dubbelop ????????? Nee dacht niet.
 
Thus, we do emphasise that this theory is local, provided that the original quantum Hamiltonian is local:
\be H^\intt_\tot=\sum_{i,j}H^\intt_{ij}\ , \eel{Hinttot.eq} where the points \(x\) associated with \(i\) and \(j\) must be neighbours only.

 A reader  thought to have found a counter example in the two-particle system. If the interaction Hamiltonian is written as a potential function \(V(x_i,\,x_j)\), of course both the quantum and the classical theory are non-local. If states with indefinite particle numbers are considered, the only known local theory is quantum field theory (QFT), where the Hamiltonian contains kinetic parts exactly of the form  \eqn{Hinttot.eq}. Indeed, the Standard Model is a QFT with all desired locality properties.
 
 Since we frequently use a lattice formulation, we do have to indicate more precisely what we mean by locality in a lattice theory. There, the best approximation to locality for the Hamiltonian is a Hamiltonian that is the sum over the entire lattice of a Hamiltonian density, \(H=\sum_{\vec x}\,\HH(\vec x)\), where \(\HH(\vec x)\) must obey 
 \be [\HH(\vec x),\,\HH(\vec y)]=0\quad\hbox{if}\quad |\vec x-\vec y|>a\ ,\eel{commutator.eq}
 where \(a\) is the lattice mesh size. The limit \(a\downarrow 0\) then reproduces locality (and causality) as it usually is formulated in QFT.
 Our classical formalism will be local in the same sense. It implies that all of the fast variables \(\vv_i\) are only allowed to affect the Hamiltonian density component(s) that sense the same operators at the coordinate \(\vec x_i\). 
 
 Thus, contradicting some claims to the contrary, we have no problem with locality. In contrast, we do have a problem with special relativity. The Hamiltonian that we start from may easily be chosen to be the one of the Standard Model, which is Lorentz covariant. However, its lattice version is in general not Lorentz covariant. Although this disease does not seem to be serious, as Lorentz invariance will naturally be recovered in the continuum limit, it will have to be a topic for discussion for the time being\fn{Lattice theory is a known and accepted method for regularising and renormalising a QFT.}.

\newsecl{About Bell's theorem}{bell.sec}
 	J.S. Bell\,\cite{Bell.ref,Bell1982.ref,Bell1987.ref} explicitly emphasised that hidden variables in general, and local hidden variables in particular, should be incompatible with quantum mechanics. To start with, there seem to be just three options. First, there may be fatal flaws in the present paper. I am sure that quite a few readers will be quite comfortable with this possibility and not look further. The author would be eager to know where they disagree with our approach; there is no fatal flaw. Small inaccuracies in the formalism would not invalidate the fundamental principle stating that the quantum formalism can be applied just as much in deterministic theories as in wave mechanics with uncertainty relations. All that was done here is turn this observation around: there is no reason to draw a sharp dividing line between quantum theories that have a deterministic foundation, such as here, and theories where this should be fundamentally impossible: we claim that deterministic theories are a dense subset of all quantum mechanical theories.
	
	Secondly, there could be a flaw in Bell's arguments. Now his derivations are clear, and it is generally agreed that, given his assumptions, his derivations are correct. The technical part of his arguments seems to be flawless. It has been argued that different number tsystems, non-commuting numbers, or other concoctions should be used, but that would be hard to defend since, whether a calculation is correct or not, should not depend on mathematical procedures and number systems employed.
	
	There is a third explanation, which seems to be somewhat more plausible than the other two: Bell did make assumptions, both concerning the nature of the hidden variables, and the nature of physical law. Bell's assumptions  are notoriously controversial. For a discussion, see the papers contained in Bell and Gao\,\cite{MaryBell.ref}. An important assumption made by Bell is ``statistical independence": the idea that due to irreproducible changes in the background, any unwanted correlation between the photons and the settings chosen should disappear. In our models, scrambling the data would require  changes in the basis elements of the cellular automaton, such that the entangled photons cannot stay in the same entangled state. The Schr\"odinger equation derived here remains the same, but not its realistic basis elements.
		
	Did Bell think of fast fluctuating hidden variables? It appears that he would have answered this himself with yes, but his assumptions seem not to include this possibility in-depth.
Usually, authors discussing Bell's theorem seem to have extremely conventional classical systems in mind, but most of them would have argued that having local hidden variables move very fast would not affect their results.

Bell did make other assumptions that this author views as suspect, see for instance\,\cite{Vervoort2013.ref}. He discusses at length the notions of causality and ``retro-causality": the settings chosen by Alice and Bob cannot affect the polarisations of the photons they observe, since ``these photons were there earlier". Can't they? In any case, this is not enough  to argue that the settings chosen by Alice and Bob must be \emph{statistically independent} of the polarisations of the photons. It is a standard calculation to find out that quantum theory does generate statistical correlations\,\cite{GtHCA.ref}. Classical theories can do the same thing. Bell derives his theorem by assuming the absence of such correlations.

It is well-known that there are statistical correlations everywhere. For instance, if there is a laptop somewhere to be found in the universe, then the odds are very high that there are more laptops in its vicinity, even space-like separated ones. In other parts of the universe there are no laptops around at all.  This is an example of a non-local correlation function. The correct way to distinguish forward from backward causality is not about statistics but about effects that would be connected by laws of nature. In QFT, a sound definition of causality can only take one form, where we cannot distinguish forward from backward: 
\begin{quote}\emph{Observable operators must commute when space-like separated}\end{quote}
(see Eq.~\eqn{commutator.eq} for the Hamoltonian density).
In quantum field theory,  statistical \emph{correlations} are expressed in the \emph{propagators} connecting space-time points. They are non-vanishing all over space and time. The \emph{commutator} of two operators vanishes completely outside the light cone.\fn{As stated earlier, special relativity is notoriously hard to obtain on a finite lattice, and at first sight, there may not be a lightcone at all. However, the cellular automaton usually does have a speed limit for signals, even if nearest neighbors can communicate directly.}

 Bell knew about this formulation of causality (``No Bell telephone") but found it to be not enough. Indeed, if only this form of causality would be allowed he would not have been able to prove his theorem. It is important to realise that the equations of motion are needed if one desires to distinguish forward from backward causality.
 
Cause and effect are statistically correlated. Thus, our option \#3 is that, by way of the `butterfly effect', even minor fluctuations in a photon wave function in the past could generate a correlation with settings Bob and Alice decide about much later as well as much earlier.

Do Alice and Bob have no `free will' then? Not in a deterministic world, it seems. Many experienced scientists have difficulties with that\,\cite{CK2008.ref}.
Physicists desire more solid arguments for the emergence of statistical correlations between the settings chosen by Alice and Bob, and the fluctuations of photons that existed at earlier times, in such a way that every hint of conspiracy can be avoided.

The notion of causality, and the demand that there should be no `retro-causality', seem to be quite plausible intuitively. It is intuitively false to allow for theories where photons in different parts of the universe `conspire' with one another. 
Let us now announce that there exists a much more powerful explanation for the apparent `conspiracy' in Bell's set-up. This, option \#4, goes as follows:

Our classical description of \emph{any} quantum system comes at a price: we \emph{first} must choose for an orthonormal basis (for the slow states \(|i\ket\)). \emph{Then} we add the fast variables, in such a way that the entire system evolves with probability distributions as described by the given Schr\"odinger equation. But how does this compare with the system we get if all states would have been transformed to a different basis? This happens when Alice or Bob change their settings.  In that case, the Schr\"odinger equation looks different, and the probability distributions are calculated in a different way. It is inevitable that the realistic states, including the initial states, in both cases are chosen  different. In Bell's set-up, it has been tacitly assumed that only one set of realistic states is accepted to be the `physically correct one'.  No, the physically correct realistic states inevitably depend on the basis chosen. The automaton allows to choose the basis, but it will have to be modified. This produces precisely the `loophole' needed to avoid Bell's conclusions.

If we now work with superpositions, such as states \(|a\ket=\alpha_1|i\ket+\alpha_2
|j\ket\) and \(|b\ket=\beta_1|i\ket +\beta_2|j\ket\), then the Born probabilities do \emph{not} apply to the inner products \(\bra a|b\ket\).\fn{This is because, if only the states \(|i\ket\) and \(|j\ket\) have an ontological meaning, then the probabilities for having a state \(|a\ket\) or a state \(|b\ket\) are meaningless.}  In the new basis, the realistic states differ from what they were before. It is here that we are not allowed to assume the photons to stay in the same entangled state. This is because the ontological theory eventually predicts only ones and zeros as outcomes. \begin{quote} \emph{If we start with realistic photons in our initial states} (photons that can only be in states either with probability one or with probability zero) \emph{then a change in Alice's and Bob's settings later in time, will force us to perform a basis transformation first.} \end{quote} We are then forced to choose how to change our realistic photons in the initial state. 
What is nice about this explanation is that we only have to reformulate our basis of states, which looks as a `conspiracy', but it isn't that. \emph{If Alice and bob's actions also fit in the deterministic equations, then there is no further need to change the basis. The deterministic model is a totally natural one.} This implies that indeed Alice and Bob have no free will, but they do not have to fear a `conspiracy plot'.

Our `change of basis' will be typically the basis transition caused by a symmetry such as a rotation. But this means that:
\begin{quote} Symmetry transformations such as rotations may transform realistic states into superpositions! \end{quote}

This would actually be an argument against our construction. However, there are ways to avoid the need for such symmetry structures: we may postulate that  basis elements are chosen as realistic states, \emph{only if these among themselves obey the same symmetry principles as the quantum system we are attempting to describe.} In the case of photons, the obvious choice is to take the vector potential fields located at well-defined space-time points. Rotating the polarisation of a photon then requires rotating the vector potential field, and this will now be an ontological transformation (the equal-time commutators for vector potential fields vanish). However, in this case the photon states themselves are not realistic: photons are energy packets of these vector fields and these do not commute.

Following this line of argument, Bell was not allowed to assume the presence of a photon with a given polarisation anywhere. Only the fields are real. Taking this new constraint into consideration, it is possible to construct a model that exactly reproduces the standard  quantum mechanical predictions for Bell's experiment. Bell would not have accepted such models because every setting chosen by Alice and Bob would require different microscopic configurations everywhere, leading to the absence of `free will'  for Alice and Bob.

In this special case, there would still be an other issue: due to lack of gauge invariance, not the vector potential but the electric and magnetic fields are well-defined, see section~\ref{model.sec}.

 \newsecl{Note added: Quantum Weirdness}{weird.sec}
  
 It may seem odd that quantum mechanics, with all its remarkable and paradoxical properties, can be explained at all using fairly mundane classical, deterministic degrees of freedom. Many investigators were confident that such simple explanations should be impossible. And yet, here we are: our model has two kinds of dynamical variables, the slow ones, \(|i\ket\), which we proclaim to be ontologically observable, and in addition fast moving variables that vaguely resemble heat baths (which they are not). The reason why, nevertheless, this system can accurately mimic quantum mechanics is that it only does so after we choose a preferred basis: the set of basis states represented by the \(|i\ket\).  
 
 All we did is to find extra variables that force the probabilities \(P_i=\bra i|i\ket\) to be such that the states \(|i\ket\) accurately obey a Schr\"odinger equation. What is not allowed is to change the basis for these states \(|i\ket\) half-way an experiment. We should keep in mind that Born's rule for the probabilities does not apply when one superposition is compared with another superposition; it only applies if a superposition is compared to realistic states.
 
 Practically all examples of counter intuitive quantum features such as the EPR-Bell experiments, the GHZ state\,\cite{GHZ.ref}, and similar
constructions, require as a crucial step the freedom of one or more of the observers, be it Alice, Bob, Cecile, Dave, \dots, to \emph{change their basis of states}, for instance by rotating their polarisation devices.  In our set-up, the fast variables change completely after any such change, regardless how small. The original state that these set-ups start from is entangled some way, and any change of basis states gives  the initial particles a completely different probabilistic distribution -- as if some signal went back to the past to instruct this initial state. Of course, the actual configuration of classical variables in our set-up does not allow for such a signal, which means that, in our models, the observers do not have the `free will' to rotate their detectors without any modifications of the (quantum or classical) states in the past.

 \newsecl{Model building}{model.sec}
 One reader suggested that our theory can be compared with Bohm's pilot theory\,\cite{Bohm.ref}. A big difference is that Bohm's theory appears to hinge on an essential statistical component: the pilot wave generates a distribution of particle positions. We claim that our approach also applies to particles that do not spread at all into statistical wave forms. Most notably, if it is our fast variable \(\vv_i\) that one would be tempted to compare with pilot waves, the contrast is that our \(\vv\) variables might be interpreted as representing the most massive virtual particles in some construction of unified theory of elementary particles.
 
 So-far, neither special relativity, nor general relativity entered in the discussion. We also ignored theories with global or local gauge symmetries. Taking these various very special symmetries of nature into account may however be tremendously important.
 Special relativity requires the replacement of simple-minded quantum mechanics by quantum field theory, QFT.  Keeping QFT discrete would be tantamount to putting the theory on a lattice. In the classical case one then obtains a \emph{cellular automaton}\,\cite{GtHCA.ref},\,\cite{Zuse.ref}--\cite{Wolfram-2020.ref}. Turning this in a quantum theory requires procedures as described in this paper\,\cite{Wetterich.ref,GtHdetmath.ref,GtHonto.ref}.  %page 12
 Now QFT is a quantum theory just as any other one, except that it may possess Lorentz invariance as a symmetry. Can we take Lorentz invariance into account? 
 
 Lorentz invariance requires locality in quantum fields, but locality does not impose any difficulty at all. Our fast fluctuating variables, \(\vv_i\), may very well be strictly localised in points of space and time.\fn{In a previous version of this paper, the fast variables \(\vv_i\)  were associated directly to the states \(|\psi_i\ket \) of the slow variable. This would work for a one-particle case, but should not be done in more complex quantum systems, particularly if we wish to recover locality. The index \(i\) for the variables \(\vv_i\) must be associated to  coordinates \(\vec x\). I thank the referee for spotting this inaccuracy.}   The problem with the Lorentz group, however, is that it is not compact. By applying Lorentz transformations successively, one can reach ever stronger Lorentz boosts. Where is the end? Is there an end?
 
We may attempt to construct a theory with Lorentz invariance containing fast fluctuating variables: when strongly boosted, any particle system may then be transformed into a highly energetic one. These can also play the role of fast variables. Our problem may be the converse: our variables \(\vv_i\) may move very fast but they were postulated only to come in a finite number of states. A Lorentz invariant theory will necessarily have an infinite number of states, as dictated by the infinite Lorentz group. Is there a way to `compactify' the Lorentz group?
 
 There is a mathematical trick that is very useful in QFT: the Wick rotation\cite{Symanzik1968}. This transforms real time into imaginary time, and, miraculously, turns the Lorentz group into the group of rotations in a 4 dimensional, Euclidean space. This group is compact, so that our problem disappears. But now we have no unitary evolution law anymore. Instead, our system now describes interactions in a stochastic system at equilibrium.

There is a \emph{physical} argument telling us that there will be a limit to the Lorentz boosts: gravity. In ordinary physical circumstances, there is only one state with vanishing energy: the vacuum. All other states contain a positive amount of energy. If we Lorentz-boost such a state, we get particles with a large amount of energy as well as momentum. These energies and momenta generate gravitational fields, and with those, curvature in space and time. Consequently, The Lorentz boosts, which lie at the center of \emph{special relativity theory,} can be terminated by 
 the theory of \emph{general relativity}. This is caused by the fact that there will be limits to the amount of space-time curvature that we may be able to handle in our system.
 
 The present paper suggests that we should not worry too much about quantum mechanics when this problem arises, but we should try to formulate classical evolution laws where these features are taken into account. This is also very difficult, but at least it may seem to be a more manageable problem.
 
 Physics is about making accurate models of the real world around us. This paper encourages us to continue doing this. The most advanced model we have today is the Standard Model. It consists of a small set of fundamental fields, which interact in ways where we need a few dozen of freely adjustable parameters, the constants of nature. A fundamental problem today is to speculate what the physical origin of these numbers might be. What one can notice from the results of this paper is, that we cannot have just any set of real numbers for these interaction parameters. Finiteness of the lattice of fast fluctuating parameters would suggest that, if only we could guess exactly what the fast moving variables are, we should be able to derive all interactions in terms of simple, rational coefficients. Thus, a prudent prediction might be made: 
 \begin{quote}
 \emph{All interaction parameters for the fundamental particles should become calculable in terms of simple, rational coefficients}. \end{quote}
 Needless to say that we are unable to pin down more precisely how to predict these values today, but we can say, for instance, that a smooth space-time dependence of the interaction  parameters, as is being speculated about by some investigators, is not possible in the framework of this paper.
 
 \newsecl{Concluding remarks}{conc.sec}
 Every cellular automaton allows for a description in terms of a quantum Hamiltonian, which reproduces the evolution of all cells with infinite perfection. Conversely, every quantum system can be approximated by an Hamiltonian derived from a cellular automaton. Since there appears to be a continuum of distinct quantum field theories but only a denumerable number of cellular autumaton rules, this latter mapping cannot always be perfect, but, if a limit on the time resolution is allowed, the deviations (in terms of energy eigenvalues) can be made arbitrarily small. In particular, we showed how, in principle, to construct a cellular automaton to match the Standard Model of the elementary particles.
 
 This inspires us to formulate our physical theory for the interpretation of quantum mechanics:
 \begin{center} \emph{Our universe is a cellular automaton.} \end{center}
 A special future of this hypothesis is that it can be phrased with infinite perfection, and this entails that the universe may have begun with a single, fundamental state, and will continue to be in realistic states forever. This actually allows one to counter the persistent critical objection that the cellular automaton does not seem to describe the evolution of superimposed states correctly:
 
  \emph{This does not matter; the universe never is in a quantum superposition of CA states!}
 
 This sounds like a fundamental departure from Copenhagen, but it is easy to defend. The reason why nevertheless, interference patterns and superposition phenomena are observed, is that such phenomena are only observed if an experiment is repeated many times. What happens when an experiment is repeated many times is described correctly by our Hamiltonian.
 
 This now leaves a lot to be done: \emph{Now do the calculation and find the rules for the world's automaton!} Of course, this is extremely difficult. We still haven't quite understood symmertries such as Lorentz invariance, and, certainly, gravity has not been understood. A good feature of our theory is that, now, we only need to investigate realistic theories for gravity, which may be conceptually much easier to do than keeping everything `quantum'. Why not try?
 \\[10pt]
 The author thanks A. Schwarz and C.~Wetterich for interesting discussions. We also had constructive, though sometimes  fierce,   discussions in weblogs with Ron Maimon, Mitchell Porter, Alan Rominger, Manuel Morales, Lubo\u{s} Motl, and others. Among them were, and no doubt still are, pertinacious nay-sayers.

 \end{document}